\documentstyle[12pt]{article}

\advance\textheight by \topskip

\begin{document}
%
\hfill
NIKHEF-H/93-26

\hfill
To appear in Phys. Lett. B

\begin{center}
{\huge \bf The
$\alpha_{s}^3$ correction to $\Gamma (Z^{0}\rightarrow hadrons)$
 }\\ [8mm]
S.A. Larin\footnote{On leave from the Institute for Nuclear
Research (INR) of the Russian Academy of Sciences, Moscow 117312.},
T. van Ritbergen, J.A.M. Vermaseren \\ [3mm]
NIKHEF-H, P.O. Box 41882, \\ 1009 DB, Amsterdam \\
\end{center}
\begin{abstract}
We present the $\alpha_{s}^3$ correction to the  $Z^{0}$ decay rate
into hadrons in the limit $m_{top} >> m_{Z}$.
\end{abstract}

\section{Introduction}
Precision measurements of the $Z^{0}$ decay rate into hadrons at LEP
\cite{lep1} provide precise means to extract the QCD coupling constant
from experiment. The $\alpha_{s}^3$
approximation to
the $Z^{0}$ decay rate into hadrons
is important for an accurate determination of $\alpha_{s}$.
The hadronic $Z^{0}$ decay rate is a sum of vector and
axial-vector contributions of which
the vector contribution is known
to order $\alpha_{s}^3$ from the calculation of $\sigma_{tot}(e^{+}e^{-}
\rightarrow \gamma
\rightarrow hadrons)$ \cite{gorishny2}.
The correctness of this calculation is strongly supported by
\cite{broadhurst2} where the non-trivial connection between the result
\cite{gorishny2} and the $\alpha_{s}^3$ approximation \cite{gls}
 to deep inelastic sum rules was established.
The axial-vector part of the hadronic $Z^{0}$ decay rate was calculated
to order $\alpha_{s}^2$ in \cite{kniehl1} and confirmed in \cite{chetyrkin1}
where renormalization group improvements were made.
The $Z^{0}$ decay into 3 gluons in order  $\alpha_{s}^3$
has been calculated in \cite{bij1}.
In this paper we present the total $\alpha_{s}^3$ correction to the
$Z^{0}$ decay rate into hadrons by calculating the axial-vector part
in  order $\alpha_{s}^3$ in the leading order of a large top mass
expansion.

\section{Preliminaries}
For the $Z^{0}$ decay rate into hadrons,
the quantity to be determined is the squared matrix element summed over
all final hadronic states.
One can express this quantity as the imaginary part of a
current correlator in the standard way
\begin{equation} \sum_{h} <0| J^{\mu} |h> <h| J^{\nu}|0> \hspace{.15cm} =
\hspace{.05cm} 2 Im \Pi^{\mu \nu} ,\end{equation}
\begin{equation} \label{correlator}
\Pi^{\mu \nu} = i \int d^{4}z \mbox{\bf e}^{i q\cdot z}
<0| T( J^{\nu}(z) J^{\mu}(0) ) |0> \hspace{.15cm}= -g^{\mu \nu} \Pi_{1}(q^{2})
-q^{\mu}q^{\nu}\Pi_{2}(q^{2}). \end{equation}
Here $ J^{\mu}=\frac{g}{4 c_{W}}\sum_{q=1}^{6} \left[
(2I^{(3)}_{q}-4Q_{q}s_{W}^{2}
) \overline{q} \gamma^{\mu} q+2I^{(3)}_{q} \overline{q} \gamma^{\mu}\gamma_{5}
q
\right] $ is the neutral weak quark current coupled to the $Z^{0}$ bozon in the
Lagrangian of the Standard Model.
The hadronic $Z^{0}$ decay width is expressed as
\begin{equation}
 \Gamma_{had} \equiv \Gamma_{had}^{V} + \Gamma_{had}^{A}
= \frac{1}{m_{Z}}Im\Pi_{1}(m^{2}_{Z}+ i\hspace{.03cm} 0^{+})
\end{equation}
with the indicated decomposition into vector and axial-vector parts imposed
by the structure of the neutral current.

Throughout this paper we use dimensional regularization \cite{dimreg}
and the standard modification of the minimal subtraction scheme
\cite{ms}, the $\overline{MS}$ scheme \cite{msbar}.
For the treatment of the $\gamma_{5}$
matrix in dimensional regularization we use the technique  described
in \cite{larin2} which is based on the original definition of $\gamma_{5}$
in \cite{dimreg}. We work in the approximation of 5 massless flavors
and the top quark mass large compared to the $Z^{0}$ mass.
We should stress that the top quark does not
decouple \cite{nodecouple} from the axial-vector part due to diagrams
of the axial anomaly type.

The $\alpha_{s}^3$ approximation \cite{gorishny2} for the
vector part in effective QCD with 5 active massless quark flavors
in the $\overline{MS}$ scheme reads
\[
\Gamma_{had}^{V} = \frac{G_{F}m_{Z}^{3}}{8\sqrt{2} \pi} \sum_{q=1}^{5}
(2I^{(3)}_{q}-4Q_{q}s_{W}^{2})^{2} \left[ 1+\frac{\alpha_{s}}{\pi}
+1.40923 \left( \frac{\alpha_{s}}{\pi} \right)^{2}
-12.76706 \left( \frac{\alpha_{s}}{\pi} \right)^{3} \right] + \]
\begin{equation} \label{gammavec}
 + \frac{G_{F}m_{Z}^{3}}{8\sqrt{2} \pi}
  \left( \sum_{q=1}^{5}(2I^{(3)}_{q}-4Q_{q}s_{W}^{2}) \right)^{2}
  \left[ -.41318 \left( \frac{\alpha_{s}}{\pi} \right)^{3} \right]
\end{equation}
with the Fermi constant $G_{F} = \frac{g^{2}\sqrt{2}}{8 c_{W}^{2} m_{Z}^{2} }$.
Here $\alpha_{s}=\alpha_{s}^{(5)}(m_{Z})$ is the coupling constant in
effective QCD with 5 active flavors (the expression for $\alpha_{s}^{(5)}$
 in NNL approximation can be found e.g. in \cite{gorishny2}).

It is convenient to split the axial-vector contribution in a
non-singlet and a singlet part
\begin{equation}
\Gamma_{had}^{A} = \Gamma_{had}^{A,NS} + \Gamma_{had}^{A,S}.
\end{equation}
The non-singlet part
comes from Feynman diagrams where both axial vertices are located in one
fermion loop.
The non-singlet part can be reduced to the vector case by using the
effective anticommutation property of the $\gamma_{5}$ matrix in the
prescription that we use. The result
in the effective theory with 5 active massless quark flavors is
\begin{equation} \label{gammaaxialns}
\Gamma_{had}^{A,NS} = \frac{G_{F}m_{Z}^{3}}{8\sqrt{2} \pi} \sum_{q=1}^{5}
(2I^{(3)}_{q})^{2} \left[ 1+\frac{\alpha_{s}}{\pi}
+1.40923 \left( \frac{\alpha_{s}}{\pi} \right)^{2}
-12.76706 \left( \frac{\alpha_{s}}{\pi} \right)^{3} \right] .
\end{equation}

\pagebreak
\section{The calculation}
 The singlet contribution $\Gamma_{had}^{A,S}$
comes from diagrams where each axial vertex
is located in a separate fermion loop.
The 3-loop and 4-loop singlet diagrams that we need to calculate
are drawn in figure 1.
 \vspace{9cm}
\[ \parbox{11.5cm}{Figure 1. The symbol $\otimes$ is used to indicate an axial
vertex.
 } \] \\
In the Standard Model quarks in a weak doublet couple with opposite
sign to the $Z^0$ bozon in the axial part of the neutral current.
That is why the contributions from light doublets add up to zero in the
massless
limit for singlet diagrams. The only non-zero contribution comes from
the top-bottom doublet due to
the large mass difference between top and bottom quarks.
The massless diagrams (i.e. without top quark loops) were calculated
with techniques similar to the ones that were used for the calculation of
$\sigma_{tot}(e^{+}e^{-} \rightarrow \gamma \rightarrow hadrons)$
\cite{gorishny2}.
The $R^*$-operation \cite{rstar} which subtracts both ultraviolet and
infrared divergences was a necessary part of these techniques.
For the calculation of massive diagrams
(i.e. with one or more top quark loops) we applied an asymptotic expansion
in a large top mass $m_{t}$.
The theory of Euclidean asymptotic expansions was developed in
\cite{tkachov1,chetyrkin2}.
For the large mass expansion of the diagrams we use the
techniques developed in \cite{gorishny1,gorishny3}.
For the actual calculation we relied heavily on the symbolic
manipulation program FORM \cite{form}.
 Massless diagrams were calculated with the
help of the package MINCER \cite{mincer1}. Massive diagrams
were calculated with a specially designed package for massive vacuum
3-loop diagrams which uses algorithms from ref. \cite{broadhurst1}.
Both packages are based on the `integration by parts' method \cite{ibp}.
Details of the calculation will be given in a longer paper where we will
also present the higher order terms in the large mass expansion.
The computations were done in an arbitrary covariant gauge. The cancellation
of the gauge dependence in physical quantities gives a good check
of the results.
The result of our calculation in the leading approximation of the large
$m_{t}$ expansion for 6 flavors is
\begin{equation} \label{main6fl}
\Gamma_{had}^{A,S} = \frac{G_{F}m_{Z}^{3}}{8\sqrt{2} \pi}
\left[ d_{2} \left( \frac{\alpha_{s}^{(6)}}{\pi} \right)^{2}
 + d_{3} \left( \frac{\alpha_{s}^{(6)}}{\pi} \right)^{3} \right] ,
\end{equation}

$ d_{2} = T_{F}^2 D \left( - \frac{37}{24} +
 \frac{1}{2}ln(\frac{m_{Z}^2}{m_{t}^2}) \right)$,

$ d_{3} = N_{f} T_{F}^3 D \left( \frac{25}{36} - \frac{1}{18}\pi^2 +
 \frac{1}{9} ln(\frac{\mu^2}{m_{t}^2})
 - \frac{1}{6} ln^{2}(\frac{\mu^2}{m_{t}^2}) - \frac{11}{12}
 ln(\frac{m_{Z}^2}{\mu^2}) + \frac{1}{6} ln^{2}(\frac{m_{Z}^2}{\mu^2})
 \right) $

\hspace{.5cm}$+ C_{A}T_{F}^2 D \left(  - \frac{215}{48} - \frac{1}{2}\zeta_{3}
 + \frac{11}{72} \pi^2
+ \frac{19}{36} ln(\frac{\mu^2}{m_{t}^2})
+ \frac{11}{24} ln^2 (\frac{\mu^2}{m_{t}^2})
+ \frac{161}{48} ln(\frac{m_{Z}^2}{\mu^2})
- \frac{11}{24} ln^2(\frac{m_{Z}^2}{\mu^2}) \right)$

\hspace{.5cm}$ + C_{F}T_{F}^2 D \left( - \frac{3}{4} + \frac{3}{2} \zeta_3
 - \frac{3}{4} ln(\frac{\mu^2}{m_{t}^2}) \right) +
 T_{F}^3 D \left( - \frac{157}{108} +\frac{1}{18}\pi^{2}
+ \frac{11}{12} ln(\frac{m_{Z}^2}{m_{t}^2})
- \frac{1}{6} ln^{2} (\frac{m_{Z}^2}{m_{t}^2}) \right). $
\\ \\
Here $m_t\equiv m_t(\mu)$ is the running top mass in the $\overline{MS}$
scheme.
$\alpha_{s}^{(6)}(\mu) = \frac{g^2}{4\pi}$ is the coupling
constant in QCD with 6 flavors,
 $C_{F} = \frac{4}{3}$ and $C_{A}= 3$ are the Casimir operators of the
fundamental and
adjoint representation of the color group $SU(3)$, $D = 8$ is the dimension
of the Lie algebra, $T_{F} = \frac{1}{2}$ is the
trace normalization of the fundamental representation,
 $N_{f} = 6$ is the number of
quark flavors, $\zeta$ is the Riemann zeta-function. Contributions
with $\pi^2$ originate
from terms containing $ln^3 (-m_{Z}^2-i 0^{+}) = (ln(m_{Z}^2) - i \pi )^3$
when one takes the imaginary part of the correlator.
Results for individual diagrams contain $\zeta_{2}$,
$\zeta_{4}$ and $\zeta_{5}$ but these contributions add up to zero in the
total result.
Note that the $\alpha_{s}^2$ order agrees with the known
calculation \cite{kniehl1}.

The coefficients of the logarithms are in agreement with the required
renormalization group invariance of the physical quantity
\begin{equation} \mu^{2} \frac{d}{d \mu^{2}} \Gamma_{had}^{A,S}
 (\mu^{2},\alpha_{s}(\mu),m_{t} (\mu) ) = 0.
\end{equation}
Of course the true physical quantity is $\Gamma_{had}$.
 But from a theoretical point
of view the singlet part $\Gamma_{had}^{A,S}$ is renormalized independently
of the non-singlet part and is therefore renormalization
 group invariant by itself.

The $\alpha_{s}^{3}$ approximation for the vector part
and the axial non-singlet part are calculated in
effective QCD with 5 active massless flavors where
the top quark decouples. Although the top quark does not decouple
from the singlet axial part we should convert it to an expression in
the effective theory to give a consistent total $\Gamma_{had}$.
The QCD decoupling relations in two loop approximation in the $\overline{MS}$
scheme were calculated
in \cite{decouple1,decouple2}. The connection between the full coupling
constant $\alpha_{s}^{(6)}$ and the coupling constant of effective
QCD with 5 active flavors reads
\begin{equation} \label{alphaeff}
 \frac{\alpha_{s}^{(6)}(\mu)}{\pi} = \frac{\alpha_{s}^{(5)}(\mu)}{\pi}
 +\left( \frac{\alpha_{s}^{(5)}(\mu)}{\pi} \right)^{2} \frac{T_{F}}{3}
ln(\frac{\mu^2}{m_{t}^2}) + O(\alpha_{s}^{3}) .
\end{equation}
Substitution of this expression in (\ref{main6fl}) gives the expression for
 effective QCD
\[
\Gamma_{had}^{A,S} = \frac{G_{F}m_{Z}^{3}}{8\sqrt{2} \pi}
\left[ \left( \frac{\alpha_{s}}{\pi} \right)^{2} \left(
- \frac{37}{12} + ln (\frac{m_{Z}^2}{m_{t}^{2}})
\right) \right. + \]
\begin{equation} \label{main5fl} +
\left.
\left( \frac{\alpha_{s}}{\pi} \right)^{3} \left(
- \frac{5651}{216} + \zeta_{3} + \frac{23}{36}\pi^2
+ \frac{31}{18} ln(\frac{\mu^2}{m_{t}^2})
+ \frac{373}{24} ln(\frac{m_{Z}^{2}}{\mu^2})
+ \frac{23}{12}  ln^2 (\frac{\mu^2}{m_{t}^2})
 - \frac{23}{12} ln^2 (\frac{m_{Z}^2}{\mu^2})
\right) \right]
\end{equation}
where $\alpha_{s} = \alpha_{s}^{(5)}(\mu)$ is the coupling constant in
effective QCD with 5 flavors.
The $\alpha_{s}^3$ term drastically diminishes the
$\mu$-dependence of eq.(\ref{main5fl}) and makes it stable in a wide interval
around $\mu\approx m_Z$.

Summing equations (\ref{gammavec}), (\ref{gammaaxialns}) and (\ref{main5fl})
and putting the renormalization scale $\mu$ in the standard way equal to
$m_{Z}$
we get the hadronic $Z^{0}$ decay width (in the
approximation of 5 massless quarks and a heavy top quark,
in the leading order of the large $m_{t}$ expansion)
\[
\Gamma_{had} = \frac{G_{F}m_{Z}^{3}}{8\sqrt{2} \pi} \sum_{q=1}^{5}
(2I^{(3)}_{q}-4Q_{q}s_{W}^{2})^{2} \left[ 1+\frac{\alpha_{s}}{\pi}
+1.40923 \left( \frac{\alpha_{s}}{\pi} \right)^{2}
-12.76706 \left( \frac{\alpha_{s}}{\pi} \right)^{3} \right] +
\] \[
 + \frac{G_{F}m_{Z}^{3}}{8\sqrt{2} \pi}
  \left( \sum_{q=1}^{5}(2I^{(3)}_{q}-4Q_{q}s_{W}^{2}) \right)^{2}
  \left[ -.41318 \left( \frac{\alpha_{s}}{\pi} \right)^{3} \right] +
\] \[
+ \frac{G_{F}m_{Z}^{3}}{8\sqrt{2} \pi} \sum_{q=1}^{5}
(2I^{(3)}_{q})^{2} \left[ 1+\frac{\alpha_{s}}{\pi}
+1.40923 \left( \frac{\alpha_{s}}{\pi} \right)^{2}
-12.76706 \left( \frac{\alpha_{s}}{\pi} \right)^{3} \right] +
\] \begin{equation}\label{eq11}
+ \frac{G_{F}m_{Z}^{3}}{8\sqrt{2} \pi}
\left[ \left( \frac{\alpha_{s}}{\pi} \right)^{2} \left(
- \frac{37}{12} + ln (\frac{m_{Z}^2}{m_{t}^{2}})
\right) +
\left( \frac{\alpha_{s}}{\pi} \right)^{3} \left( -18.65440
+ \frac{31}{18} ln(\frac{m_{Z}^2}{m_{t}^2})
+ \frac{23}{12}  ln^2 (\frac{m_{Z}^2}{m_{t}^2})
\right) \right] .
\end{equation}
Here and below $m_t\equiv m_t(m_Z)$ is the $\overline{MS}$ top mass at the
scale $m_Z$.
One may relate it to the pole mass through the expression
$m_{t}(m_Z)=m_{pole}\left[1-\frac{\alpha_s(m_Z)}{\pi} \left(
ln(\frac{m_Z^2}{m_{pole}^2})+\frac{4}{3}\right)+O(\alpha_s^2)\right]$
which is known in the NNL approximation \cite{polemass}
or relate it to $m_t(m_t)$ through the expression
$m_{t}(m_Z)=m_t(m_t)\left[ 1-\frac{\alpha_s(m_Z)}{\pi}
ln(\frac{m_Z^2}{m_t^2(m_t)})+O(\alpha_s^2)\right]$.
This would correspondingly modify the coefficients of the $\alpha_{s}^{3}$
term.

\newpage
Substitution of the values of the physical parameters into eq.(\ref{eq11})
from \cite{properties} gives (in the $\overline{MS}$ scheme)
\[
\Gamma_{had}(GeV) = 1.671 \left[ 1+\left( \frac{\alpha_{s}}{\pi} \right)
+ \left( \frac{\alpha_{s}}{\pi} \right)^2 \left( .9502
        +.1489 \hspace{.1cm} ln (\frac{m_{Z}^2}{m_{t}^{2}}) \right) \right. +
 \hspace{2cm} \]
\begin{equation}  \hspace{3cm} +
 \left. \left( \frac{\alpha_{s}}{\pi} \right)^3 \left( - 15.650
       + .2564 \hspace{.1cm} ln (\frac{m_{Z}^2}{m_{t}^{2}})
       + .2853 \hspace{.1cm} ln^2 (\frac{m_{Z}^2}{m_{t}^{2}}) \right) \right] .
\end{equation}
Note that the logarithms in the $\alpha_{s}^3$ term tend to cancel each other
for $m_{t}(m_Z) \approx 140 GeV$. For this value of $m_{t}$ the
calculated
order $\alpha_{s}^3$ singlet contribution to $\Gamma_{had} $ adds about
$20\%$ to the previously
known sum of vector and axial non-singlet contributions of order
$\alpha_{s}^3$.

{}From eq.(\ref{eq11}) one can also obtain the $\alpha_{s}^{3}$
approximation to
$\sigma_{tot}(e^{+}e^{-}\rightarrow \gamma,Z^{0} \rightarrow hadrons)$
in the energy range below the top quark threshold.

\section{Acknowledgements}
We are grateful to M. Veltman for helpful discussions. We gratefully
acknowledge the remarks of D.J. Broadhurst, K.G. Chetyrkin, A.I. Davydychev
 and A.L. Kataev
on an early version of our paper.
We want to thank the CAN foundation for the use of their computers.
\\ \\
{ \bf Note added.} We are grateful to  K.G. Chetyrkin
for informing us about a discrepancy in the coefficient of $\zeta_{3}$
between our early result and a result by K.G. Chetyrkin,
 J.H. K\"uhn and O.V. Tarasov before their publication.
This information helped us to correct this coefficient.
\\

\end{document}